\input epsf.tex
\documentclass[12pt]{article}
\usepackage{graphicx}
\usepackage{psfig,picinpar,floatflt,amssymb,epsf}
\csname @addtoreset\endcsname{equation}{section}

\def\bseq{\begin{subequation}}  
\def\eseq{\end{subequation}}
\def\bsea{\begin{subeqnarray}}  
\def\esea{\end{subeqnarray}}

\newcommand{\bbox}{\lower.2ex\hbox{$\Box$}}

\newcommand{\beq}{\begin{equation}}
\newcommand{\eeq}{\end{equation}}
\newcommand{\bea}{\begin{eqnarray}}
\newcommand{\eea}{\end{eqnarray}}
\newcommand{\ena}{\end{eqnarray}}

\newcommand {\non}{\nonumber}

\renewcommand{\a}{\alpha}

\renewcommand{\d}{\delta}
\renewcommand{\th}{\theta}
\newcommand{\thb}{\bar{\theta}}

\newcommand{\pa}{\partial}
\newcommand{\g}{\gamma}
\newcommand{\G}{\Gamma}

\newcommand{\e}{\epsilon}
\newcommand{\z}{\zeta}

\newcommand{\m}{\mu}

\newcommand{\n}{\nu}

\newcommand{\s}{\sigma}

\newcommand{\Db}{\bar{D}}

\newcommand{\Phib}{\bar{\Phi}}

\begin{document}

\begin{titlepage}
{\hbox to\hsize{July  2001 \hfill
{Bicocca--FT--01--19}}}
{\hbox to\hsize{${~}$ \hfill
{IFUM--691--FT}}}
\begin{center}
\vglue .06in
{\Large\bf Superspace approach to anomalous dimensions \\ in ${\cal N}=4$ SYM}
\\[.45in]
Silvia Penati\footnote{silvia.penati@mib.infn.it}\\
{\it Dipartimento di Fisica, Universit\`a degli studi di
Milano-Bicocca\\ 
and INFN, Sezione di Milano, piazza della Scienza 3, I-20126 Milano, 
Italy}
\\
[.2in]
Alberto Santambrogio\footnote{alberto.santambrogio@mi.infn.it}\\
{\it Dipartimento di Fisica, Universit\`a degli studi di Milano \\
and INFN, Sezione di Milano, via Celoria 16, I-20133 Milano, Italy}
\\[.5in]

{\bf ABSTRACT}\\[.0015in]
\end{center}

In a ${\cal N}=1$ superspace setup and using dimensional regularization, 
we give a general and simple
prescription to compute anomalous dimensions of composite operators 
in ${\cal N}=4$, 
$SU(N)$ supersymmetric Yang-Mills theory, perturbatively in the coupling 
constant $g$. We show in general that anomalous dimensions are responsible 
for the appearance of higher order poles in the perturbative expansion
of the two--point function and that their lowest contribution can be read 
directly from the coefficient of the $1/\epsilon^2$ pole. As a check of our 
procedure we rederive the anomalous dimension of the Konishi superfield
at order $g^2$. We then apply this procedure
to the case of the double trace, dimension 4, superfield in the ${\bf 20}$ of 
$SU(4)$ recently considered in the literature. We find that its anomalous 
dimension vanishes for all $N$ in agreement with previous results. 

\vskip 30pt
${~~~}$ \newline

\end{titlepage}

\section{Introduction}

Recently there has been a renewal of interest in four dimensional conformal 
field theories. In particular, new results are now available in
the context of ${\cal N}=4$ $SU(N)$ supersymmetric Yang-Mills theory, mainly
because of its connection with the AdS/CFT correspondence \cite{maldacena}.
The existence of non-renormalization theorems for the class of the 
so-called Chiral Primary Operators (CPO) has been suggested on the basis 
of the correspondence \cite{corr} and successively proven at perturbative 
and nonperturbative level:
these operators have no anomalous dimension and their two and 
three point functions are protected from getting quantum corrections
\cite{tutti,PSZ1,PSZ2}.

The operators of the theory have been recently classified 
according to the behavior of their anomalous dimensions in the strong 
coupling regime and in the large-$N$ limit \cite{AFP1}. In this contest it is
important to investigate anomalous dimensions also in the perturbative
regime and for any value of $N$. 

The anomalous dimension of a composite operator $\cal{O}$ can be determined 
by the renormalization of the source term $J\cal{O}$ 
appearing in the generating functional of correlation functions with 
insertions of the composite operator. Performing a quantum--background
splitting, one then needs evaluate order by order Feynman diagrams with
the external background structure $J\cal{O}$. However, in the case of
theories with a rich spectrum of composite operators,  this procedure 
is often not very convenient from a technical point of view. This is
certainly the case of ${\cal N}=4$ $SU(N)$ supersymmetric Yang--Mills
theory when considering operators that are not flavor or colour singlets. 
In these cases it can be more efficient to infer anomalous dimensions 
from the divergences of Green's functions \cite{AGJ,AFP1,BKRSanom}.
 
In this paper we develop a simple and direct method to compute perturbatively
anomalous dimensions of composite operators from their two--point function. 
The method is general and can be applied to any superconformal field theory
in four dimensions. 
Given a composite operator, we compute its two--point function 
order by order in perturbation theory. We use dimensional regularization
to deal with UV divergences which then appear as poles in the regularization
parameter $\e$. The conformal invariance of the theory constrains the result 
to have at most simple pole divergences which are associated to the short
distance singularity of the correlation function. 
However, we show that in general a nontrivial renormalization
for the operator at hand (i.e. the presence of nonvanishing anomalous 
dimensions) is responsible for the appearance of higher order poles in the 
perturbative expansion of the two-point function, much like it produces
logarithmic contributions in a point--splitting regularization 
\cite{BKRSanom,log}.
The search for higher order poles in the two--point correlators represents 
a simple but strong criterium to select operators which are subject to 
renormalization. 
Moreover, we show that the anomalous dimension can be read from 
the coefficient of the $1/\epsilon^2$ pole. 

We apply our procedure to composite operators of ${\cal N}=4$ $SU(N)$ 
supersymmetric Yang--Mills theory in a ${\cal N}=1$ superspace description. 
As a check of our method we rederive the anomalous dimension of the $N=1$ 
Konishi superfield at order $g^2$ in perturbation theory, obtaining a result
in agreement with the known value \cite{AGJ,AFGJ}.

We then determine the anomalous dimension 
of the double trace, dimension 4 operator, belonging to the ${\bf 20}$ of
the R-symmetry group $SU(4)$ appearing in the product of two dimension 2 
CPO's. This operator has been found to be protected in the large-$N$ limit
\cite{AFP1,AEPS}, and its protection for $N$ 
finite has also been conjectured \cite{AEPS,AES}, despite the fact that  
the group unitarity arguments of \cite{FZ} do not prevent it from 
acquiring anomalous dimensions (see however \cite{HH}).
This unexpected result has been found in a rather indirect way, studying 
the partial 
non-renormalization of the four point functions of CPO's. In this
paper we check in a direct and simple manner the protection of this 
operator.
We find that the anomalous dimension of the double trace operator written
in terms of ${\cal N}=1$ superfields vanishes for all values of
$N$ in agreement with the previous results. 

The paper is organized as follows. In Section 2 we describe the general 
procedure that allows to compute
the anomalous dimensions of composite operators in ${\cal N}=4$ 
$SU(N)$ supersymmetric Yang-Mills theory, by using a description in terms 
of ${\cal N}=1$ superfields. In Section 3 we make use of this procedure 
to reproduce the known value of the anomalous dimension of the 
Konishi operator at order $g^2$. In Section 4 we compute the anomalous 
dimension of the double trace operator $O^{\bf (20)}_{ij}$ at order 
$g^2$ and we find zero. The meaning of our result and its 
implications are then discussed in the concluding Section 5.
We have found convenient to add two Appendices. The first one collects our 
conventions, basic rules and useful identities to perform pertubative
calculations. Appendix B describes in details the construction of the 
double trace $O^{\bf (20)}_{ij}$ operator in terms of ${\cal N}=1$ superfields.

\section{The general prescription to compute anomalous dimensions}

${\cal N}=4$ supersymmetric Yang-Mills
theory describes the dynamics of a spin-$1$ Yang-Mills vector, 
four spin-$\frac{1}{2}$ Majorana spinors and six spin-$0$ particles in the 
${\bf 6}$ of the
R--symmetry group $SU(4)$, all the fields being 
in the adjoint representation of the $SU(N)$ gauge group.

A description convenient for perturbative calculations is the 
${\cal N}=1$ superspace description where the field content of the theory is
given in terms of one real vector
superfield $V$ and three chiral superfields $\Phi^i$ containing the six
scalars organized into the ${\bf 3}\times \bf{ \bar 3}$ of $SU(3)
\subset SU(4)$. The classical action is  
(we use notations of \cite{superspace})
\bea
S
&=&\int d^8z~ {\rm Tr}\left(e^{-gV} \Phib_i e^{gV} \Phi^i\right)+
\frac{1}{2g^2}
\int d^6z~ {\rm Tr} W^\a W_\a\nonumber\\
&&+\frac{ig}{3!} {\rm Tr} \int d^6z~ \e_{ijk} \Phi^i
[\Phi^j,\Phi^k]+\frac{ig}{3!} {\rm Tr} \int d^6\bar{z}~ \e_{ijk} \Phib^i
[\Phib^j,\Phib^k]
\label{actionYM}
\eea
where $W_\a= i\Db^2(e^{-gV}D_\a e^{gV})$, and $V=V^aT^a$,
$\Phi_i=\Phi_i^a T^a$,
$T^a$ being $SU(N)$ matrices in the fundamental representation.
The theory is known to be finite and then sits at its conformal fixed
point. Possible divergences can arise only in correlators of 
composite operators. 

In Euclidean space, we introduce the generating functional 
\beq
W[J,\bar{J}]=
\int {\cal D}\Phi~{\cal D}\Phib~{\cal D}V~e^{S + \int d^8z (J {\cal O}
+ \bar{J} \bar{\cal O})}
\label{genfunc}
\eeq
for the $n$--point functions of the generically complex composite operator 
${\cal O}$ 
\beq
\langle {\cal O}(z_1) \cdots \bar{\cal O}(z_n) \rangle ~=~
\left. \frac{\d^n W}{\d J(z_1) \cdots \d \bar{J}(z_n)}\right|_{J=\bar{J}=0}
\label{defcorr}
\eeq
where $z \equiv (x,\theta, \bar{\theta})$. The perturbative evaluation
of the $n$--point funtion is equivalent to computing    
the contributions to $W[J, \bar{J}]$ at order $n$ in the sources.
In particular, for the two--point function we need evaluate
the quadratic terms 
\beq
W[J,\bar{J}]\rightarrow \int d^4x_1~d^4x_2~ d^4\theta~
J(x_1,\theta,\bar{\theta}) \, 
\frac{F(g^2,N)}{[(x_1-x_2)^2]^{\g_0 + \g}} \, 
{\cal P}(D_\a, \bar{D}_{\dot{\a}})
\bar{J}(x_2,\theta,\bar{\theta})
\label{twopoint}
\eeq
where we have defined $\g_0$ to be the naive dimensions 
of the  operator, while $\g(g)$ is the anomalous dimension ($\g(g=0) = 0$).
The particular dependence on the coordinates
is fixed by the conformal invariance of the theory, 
${\cal P}(D_\a, \bar{D}_{\dot{\a}})$ is an operatorial expression 
built up with spinorial derivatives 
and $F(g^2,N)$ is a function of the coupling constant which can
be determined perturbatively. 

To deal with possible divergences in the correlation functions we perform  
dimensional regularization by analytically continuing the theory to 
$n$ dimensions, $n=4-2\e$.  We find convenient
to work in momentum space where a generic power factor
$(x^2)^{-\n}$ is given by 
\vskip 0.1mm
\bea 
\frac{1}{(x^2)^\n} ~=~
2^{n-2\n} \pi^{\frac{n}{2}} \frac{\G(\frac{n}{2} - \n)}{\G(\n)}
\int \frac{d^n p}{(2\pi)^n} ~\frac{e^{-ipx}}{(p^2)^{\frac{n}{2}-\n}}
\qquad && \n \neq 0,-1, -2,\cdots \nonumber \\
&& \n \not= \frac{n}{2}, \frac{n}{2} +1, \cdots
\label{basicformula}
\eea
For any power $\n \geq 2$ the factor $\G(\frac{n}{2} - \n) = 
\G(-\n+2-\epsilon)$, when analitically continued to the region of positive
arguments,  develops a $1/\epsilon$ singularity 
which encodes the divergent behavior of $(x^2)^{-\n}$ at short distances.
In performing perturbative calculations in momentum space 
this is the leading behavior which is
sufficient to look for when computing two--point correlation functions
for CPO's \cite{PSZ1}.
As extensively discussed in \cite{PSZ2} finite contributions in momentum space
would correspond in $x$--space to terms proportional
to $\e$ which give rise only to contact terms.

In this framework, we now concentrate on the perturbative evaluation of the
anomalous dimensions $\g$.
According to the standard definition in quantum 
field theory, a composite operator acquires anomalous dimension 
as a consequence of a renormalization which has to be
implemented in order to cancel divergences of correlation functions 
containing insertions of that operator. The simplest way to compute
these divergences is to 
evaluate perturbatively divergent contributions to the source term 
$J {\cal O}$ in the generating functional (\ref{genfunc}).
In dimensional regularization, divergences show up as poles
$1/\epsilon^k$ and they can be cancelled by adding suitable counterterms
to the original action in (\ref{genfunc}). 
Performing minimal subtraction scheme,
one ends up with the general structure for the source term given by
\beq
\int d^8z~ J {\cal O} \rightarrow \int d^8z~ J {\cal O}
\left( 1 ~+~ \sum_{k=0}^\infty \frac{a_k(g)}{\epsilon^k} \right)
\label{counterterms}
\eeq
Defining the bare operator
\beq
{\cal O}_B ~\equiv~ {\cal O} 
\left( 1 ~+~ \sum_{k=0}^\infty \frac{a_k(g)}{\epsilon^k} \right)
~\equiv~ {\cal O} \, Z
\eeq
the anomalous dimension is given by \cite{books} 
\beq
\g ~\equiv~ - \m \frac{\pa}{\pa \mu} \log{Z} ~=~ 
\a \frac{d a_1(\a)}{d \a}
\label{andim}
\eeq
where $\m$ is the mass scale of dimensional regularization and 
$\a \equiv \frac{g^2}{4\pi}$ the effective coupling of the theory.
It is important to notice that, being the theory finite, no renormalization 
of the fundamental fields is present, so that the renormalization function
$Z$ contains only divergences arising from Feynman 
diagrams with insertions of the composite operator. 

In order to find the anomalous dimension of ${\cal O}$ 
it is therefore sufficient to determine order by order in $\a$ the simple pole
divergences of Feynman diagrams with $J$ and ${\cal O}$ on the external legs.
However, this approach can be sometimes difficult to apply, expecially
when dealing with higher dimensional composite operators which are not 
flavor or colour singlets. In this cases
it is more convenient to evaluate the anomalous dimension
by computing perturbative contributions to singlet structures constructed
from the operator at hand. The simplest singlet we can compute is the
two--point function of the operator with itself (or 
its hermitian conjugate if it complex). 
We are then going to discuss in details how anomalous dimensions 
can be easily read from the higher order poles
in the two--point function. 

We will be mainly concerned with the class of
nonderivative operators which are functions  of the fundamentals 
superfields $\Phi$, $\bar{\Phi}$ and $V$. For composite operators which
are omogenoeus polynomials of the scalar superfields the naive dimension
$\g_0$ is simply the number of powers of scalars appearing in its 
expression (remember that in 4d the fundamental
scalar superfields have dimension one, while $V$ is dimensionless).     
In $n$ dimensions, $n=4-2\e$, the naive dimension of 
the operator is continued to the value $\g_0(1-\e)$. Therefore, the power
factor appearing in (\ref{twopoint}), when using 
(\ref{basicformula}) is given by
\beq 
\frac{1}{(x^2)^{\g_0(1-\e )+\g}} ~=~
2^{4-\g_0 -2\g +2\e(\g_0-1)} \pi^{2-\e} \frac{\G(2-\g_0-\g +\e(\g_0-1))}
{\G(\g_0(1-\e) +\g)}
\int \frac{d^n p}{(2\pi)^n} ~\frac{e^{-ipx}}{(p^2)^{2-\g_0-\g+\e(\g_0-1)}}
\label{basicformula2}
\eeq
For $\g$ small, the gamma function $\G(2-\g_0-\g+\e(\g_0-1))$ is singular 
for $\e \to 0$, whenever
$\g_0$ is an integer grater or equal to 2. To extract this singularity we 
perform analytic continuation by using the standard identity 
$x\G(x) = \G(x+1)$. We write
\bea
&& \G(2-\g_0-\g+\e(\g_0-1)) ~= \non\\
&& ~~~~~\frac{1}{(2 -\g_0 - \g +\e(\g_0-1))(3 -\g_0 - \g +\e(\g_0-1)) \cdots
(-1- \g +\e(\g_0-1))} \non\\
&& ~~~~~~~~~~~\times \frac{1}{-\g + \e(\g_0-1)} \, \G(1-\g+\e(\g_0-1)) 
\label{gamma}
\eea
When taking the limit $\e \to 0$ potential divergences can arise only 
from the factor $1/(-\g +\e(\g_0-1))$ for $\g$ small. 
Keeping $\e$ finite, we expand this factor
in powers of the anomalous dimension
\beq
\frac{1}{-\g + \e(\g_0-1)} ~=~
\frac{1}{\e(\g_0-1)}  \left[ 1 + \frac{\g}{\e(\g_0-1)} 
+ O\left(\left( \frac{\g}{\e(\g_0-1)}\right)^2\right) \right]  
\label{gamma2}
\eeq
Inserting back in (\ref{basicformula2}) we can write
\bea
&& \frac{1}{(x^2)^{\g_0(1-\e)+\g}} ~\sim~ 
\pi^{2-\e} \, \frac{\G(1-\g+\e(\g_0-1))}{\G(\g_0(1-\e)+\g)} ~\times
\non\\
&&~~~~~~
\frac{4^{2-\g_0-\g+\e(\g_0-1)}}{(2-\g_0-\g+\e(\g_0-1))(3-\g_0-\g+\e(\g_0-1)) 
\cdots (-1-\g+\e(\g_0-1))(\g_0-1)}
\non\\  
&&\left[ \frac{1}{\e} ~+~ \frac{\g}{(\g_0-1)} \,
\frac{1}{\e^2} 
~+~ \frac{1}{\e} \, O\left( \frac{\g^2}{\e^2}\right) 
\right] \, \times \int \frac{d^n p}{(2\pi)^n} ~
\frac{e^{-ipx}}{(p^2)^{2-\g_0-\g +\e(\g_0-1)}} \non\\
&&~~~~~~~~~~
\label{basicformula3}
\eea
This relation immediately gives a general criteria to establish 
whether a composite operator has anomalous dimensions: The
existence of a nonvanishing anomalous dimension 
is signaled by the presence of higher order poles in the $\e$--expansion
of the two--point function. 
On the contrary, whenever the operator has $\g=0$ the two--point
function has at the most $1/\e$ poles at every order of perturbation theory.
This is in agreement with the results found \cite{PSZ1,PSZ2} in the 
perturbative evaluation of two--point functions for CPO's beyond leading 
order, and proves the more general no--go theorem according to which the
two--point correlator of any protected operator cannot diverge faster than
$1/\e$, for $\e$ small. 

In the case of nonvanishing anomalous dimensions, the identity
(\ref{basicformula3}) can be used to actually compute $\g$. 
If we write the anomalous dimension as a perturbative expansion in the
coupling constant
\beq
\g ~=~ \sum_{n=1}^{\infty} a_n (g^2)^n
\label{gammaexp}
\eeq
the expression (\ref{basicformula3}) can be organized as a double expansion
in powers of $g^2$ and $\e$. At zero order in the coupling constant the
identity (\ref{basicformula3}) gives the expected $1/\e$ divergence which 
accounts for the singular behavior of the correlation function at coincident 
points
\beq
\frac{1}{(x^2)^{\g_0(1-\e)}} ~\sim~ \frac{1}{\e} \, 
\frac{\pi^{2}}{\G(\g_0)} 
\frac{4^{2-\g_0}}{(2-\g_0)(3-\g_0) \cdots (-1)(\g_0-1)} \, 
\times \int \frac{d^4 p}{(2\pi)^4} ~
\frac{e^{-ipx}}{(p^2)^{2-\g_0}}
\eeq 
At order $g^2$ we keep only linear terms in $\g$ ($\g =
a_1 g^2$) and taking the limit $\e \to 0$,
the leading order behavior of the correlation function is given by 
\beq
\frac{1}{(x^2)^{\g_0(1-\e)+\g}} ~\sim~ \frac{\g}{\e^2} \,
\frac{\pi^{2}}{\G(\g_0)} 
\frac{4^{2-\g_0}}{(2-\g_0)(3-\g_0) \cdots (-1)(\g_0-1)^2} 
\, \times \int \frac{d^4 p}{(2\pi)^4} ~
\frac{e^{-ipx}}{(p^2)^{2-\g_0}}
\eeq 
The anomalous dimension is then given by $(\g_0-1)$ times the 
ratio of the coefficients of the $g^2$ and $g^0$ leading divergences. 

Practically, we compute all the two--point diagrams contributing to 
$W[J,\bar{J}]$ with external $J$, $\bar{J}$ sources 
up to $g^2$--order, directly in momentum space. 
We obtain in general simple pole divergent contributions at tree
level and $1/\e^2$ poles at order $g^2$. We extract the coefficients
of these two divergent terms and from their ratio we read the anomalous 
dimensions.
Possible subleading divergences arising at order $g^2$  
give nontrivial quantum corrections to the coefficient $F$ in 
(\ref{twopoint}).
  
The extension of our procedure to higher orders in $g^2$ is in principle
trivial: At every order it is sufficient to determine the coefficient 
of the double pole divergence. However, since from the $g^4$--order on, this
is expected to be a subleading divergence, one has 
to deal with possible scheme dependent contributions.

\section{Anomalous dimension of the Konishi scalar superfield}

As a check of our procedure we rederive in a manifestly supersymmetric
way the known result \cite{AGJ,AFGJ}  
for the lowest order contribution to the anomalous dimension of the Konishi 
operator. 
In ${\cal N} =1$ superspace the gauge invariant Konishi superfield for
${\cal N}=4$ SYM theory is
\beq
K = {\rm Tr} \{e^{-gV} \bar{\Phi}_i e^{gV} \Phi^i \}
\eeq
where a sum over the $SU(3)$ flavor index is meant.   
We determine its anomalous dimension using both the standard approach which 
amounts to compute quantum corrections directly to the source term in 
(\ref{genfunc}) and our procedure which requires the evaluation of 
$1/\e^2$ terms in the two--point function. 
In both cases the general strategy we follow is: we select diagrams 
contributing at order $g^2$ and compute all factors coming from combinatorics, 
colour and flavor structures. We then perform the
superspace $D$-algebra following standard techniques so reducing the 
various contributions to multi-loop momentum integrals. We evaluate the 
divergent part of these integrals by using massless dimensional 
regularization and minimal subtraction scheme 
\footnote{Notice that, even
if the theory is finite, multiploop
diagrams containing insertions of anomalous composite operators
do possess UV divergent subdiagrams.}.
We observe that gauge-fixing the classical action
requires the
introduction of corresponding Yang-Mills ghosts. However they only
couple to the vector multiplet and do not enter our specific calculation.

Useful formulae for dealing with the colour and flavor structures and to 
compute the momentum integrals are collected in appendix A.

To compute one--loop contributions to the source term $J K$ in the generating
functional we first perform quantum--background splitting and concentrate
on contributions with external background structure 
$J {\rm Tr}\{\bar{\Phi}_i \Phi^i \}$. In fact, in order to
compute the anomalous dimension of $K$ it is sufficient to consider 
contributions at zero order in the background $V$ since higher 
order contributions in $V$ will necessarily give the same answer due to 
gauge invariance. 

One loop contributions are given by diagrams in Fig. $1$ where $J$ is 
a classical real source. 
Performing the 
$D$--algebra on the diagrams $(1a)$ and $(1b)$ and keeping only potentially
divergent contributions reduces them to self--energy type integrals
\beq
\int \frac{d^nk}{k^2 (p-k)^2} 
~\sim~ \frac{1}{\e} 
\label{selfenergy}
\eeq 
where the identity (\ref{selfenergy2}) has been used to select the leading 
behavior for $\e \to 0$.
No $D$--algebra must be performed on diagrams $(c)$ and $(d)$ which
again give self--energy type contributions. 

\vskip 18pt
\noindent
\begin{minipage}{\textwidth}
\begin{center}
\includegraphics[width=0.60\textwidth]{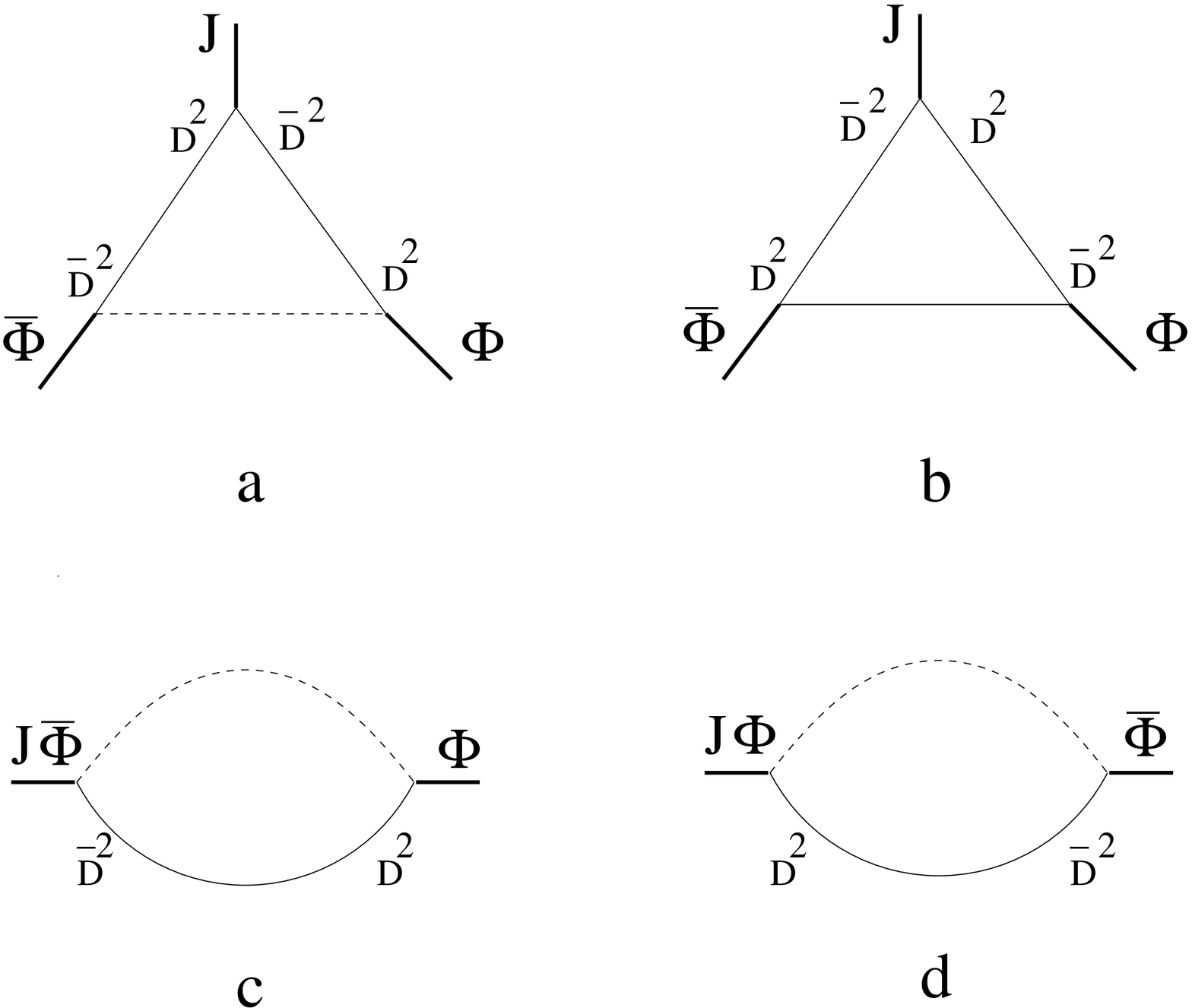}
\end{center}
\begin{center}
{\small{Figure 1:
One--loop contributions to $J {\rm Tr}\{\bar{\Phi} \Phi \}$. Dashed lines
correspond to vector propagators}}
\end{center}
\end{minipage}

\vskip 20pt

Computing
the overall coefficient from combinatorics, coefficients from vertices
and propagators and colour and flavor structures, we eventually have
\bea
&&~(1a) ~\rightarrow~ ~~\frac{2g^2N}{(4\pi)^2} \, \frac{1}{\e} 
\nonumber \\
&&~(1b) ~\rightarrow~ -\frac{4g^2N}{(4\pi)^2} \, \frac{1}{\e}
\nonumber \\
(1c) &=& (1d) ~\rightarrow~ -\frac{2g^2N}{(4\pi)^2} \, \frac{1}{\e}
\eea
These divergent contributions can be cancelled by adding a suitable 
counterterm to the source term in the generating functional  
\beq
\int d^8z \, J K ~\rightarrow~ \int d^8z \, J K \left( 1 ~+~  
\frac{3 g^2 N}{8 \pi^2} \frac{1}{\e} \right)
\eeq
Comparing with eq. (\ref{counterterms}) and using the identity (\ref{andim}) 
we finally obtain $\g = \frac{3 g^2 N}{8 \pi^2}$, as expected
\footnote{Note that there is a mismatch of a factor $2$ compared to the
result in \cite{AGJ, AFGJ} due to the fact that our coupling constant is
$\sqrt{2}$ times the one used there.}.

We now use the alternative procedure to extract the anomalous dimension 
from the two--point correlation function. 
At tree level the contribution to the two--point function $\langle 
K(z_1) K(z_2) \rangle$ is given in Fig. $2$. 
  
\vskip 18pt
\noindent
\begin{minipage}{\textwidth}
\begin{center}
\includegraphics[width=0.60\textwidth]{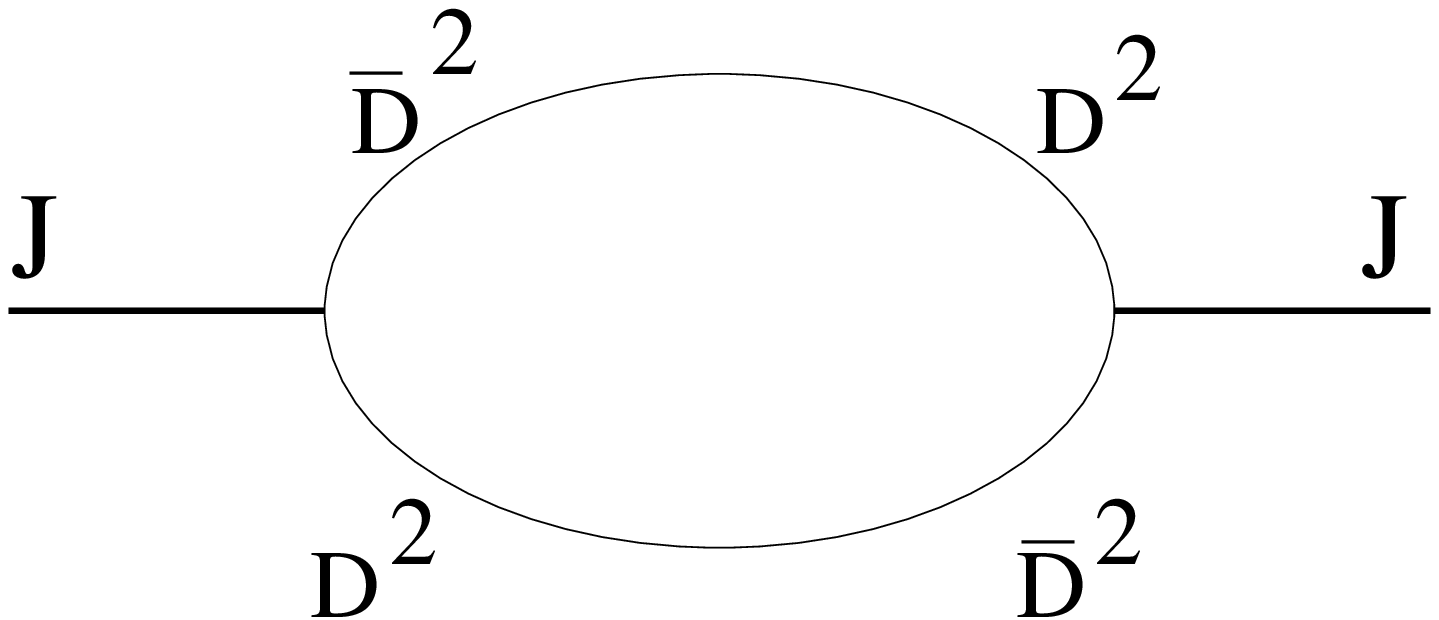}
\end{center}
\begin{center}
{\small{Figure 2: Tree level contribution to $\langle K(z_1) K(z_2) 
\rangle$ }}
\end{center}
\end{minipage}

\vskip 20pt
Performing the $D$-algebra and using eq. (\ref{selfenergy3})
we end up with two different contributions, both associated to 
self--energy type integrals (\ref{selfenergy}) but 
with $\bar{D}^2 J D^2 J$ and $\bar{D}^{\dot{\a}} J i\pa_{\a \dot{\a}} D^{\a}J$
as external source terms.
The colour and flavor structures which emerge after contraction are
\beq
{\rm Tr} (T_a T_b) {\rm Tr} (T_a T_b) \, \d^i_j \d^j_i ~=~ 3 (N^2-1)
\eeq
where the identity (\ref{norm}) have been used. Putting toghether with the 
combinatorics, the coefficients from vertices and propagators and the
result (\ref{selfenergy}) we finally obtain 
\beq
W[J]_{tree} ~\rightarrow~ \frac{1}{\e} \, \frac{3(N^2-1)}{2(4\pi)^2} 
\, \int \frac{d^4p}{(2\pi)^4} d^4 \theta \,  J(-p, \th, \thb) 
\left[\bar{D}^2 D^2 
~+~ \frac12 p_{\a \dot{\a}} \bar{D}^{\dot{\a}} D^{\a} \right] J(p,\th,\thb)    
\label{tree}
\eeq
The lowest order perturbative contribution is given by diagrams in Fig. $3$.
To draw diagrams $(3a)$ and $(3b)$ we only need the interaction vertices 
(\ref{vertices}), whereas $(3c)$ and $(3d)$ also contain the external vertex
$g J {\rm Tr} ( [ \bar{\Phi}_i , V] \Phi^i)$ from the expansion of the Konishi
operator in powers of $gV$.
 
\vskip 18pt
\noindent
\begin{minipage}{\textwidth}
\begin{center}
\includegraphics[width=0.60\textwidth]{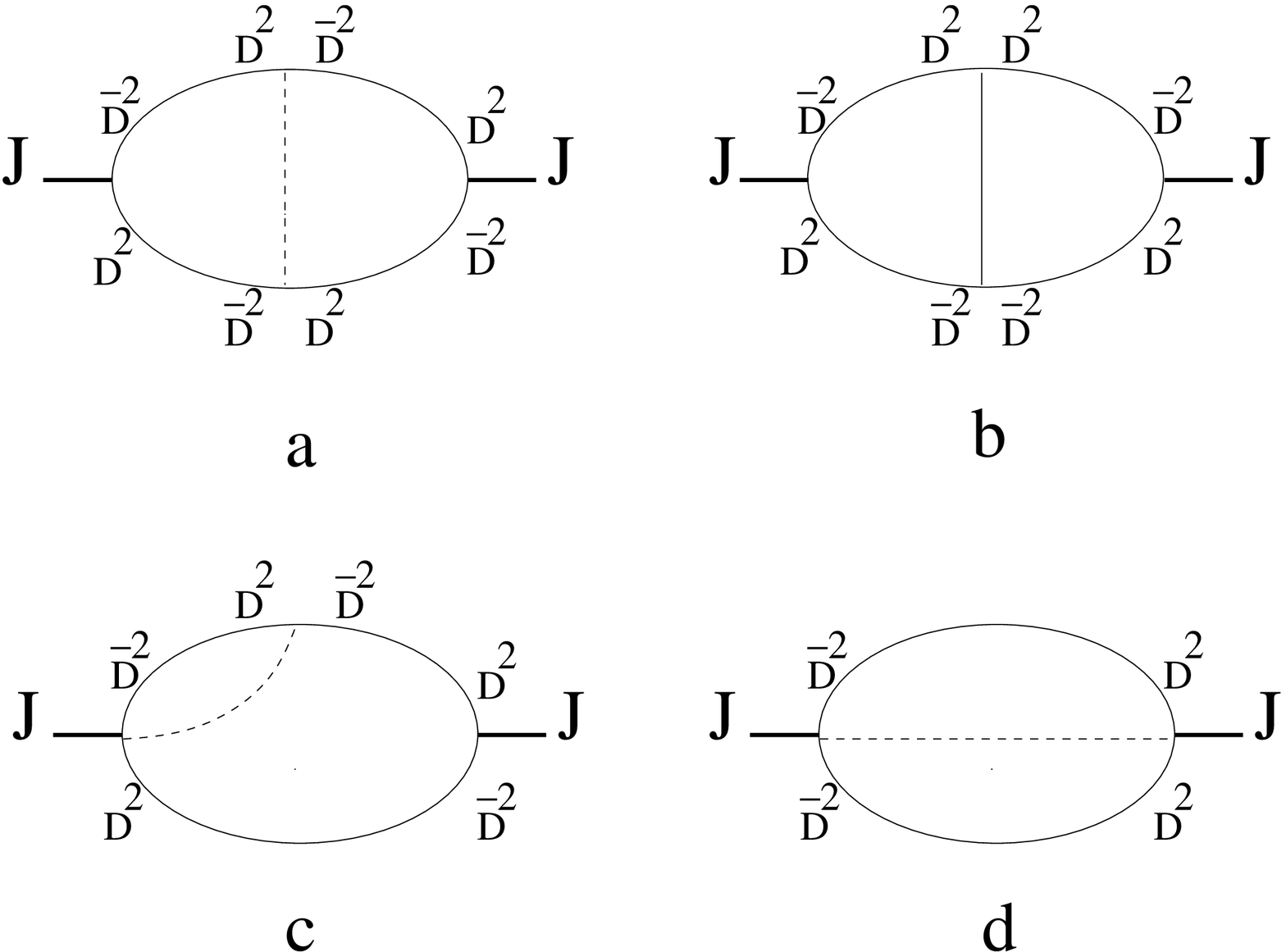}
\end{center}
\begin{center}
{\small{Figure 3:
First order contribution to $\langle K(z_1) K(z_2) \rangle$}}
\end{center}
\end{minipage}

\vskip 20pt
\noindent
We perform the $D$--algebra on the
diagrams $(3a)$, $(3b)$ and $(3c)$ (for the diagram $(3d)$ no $D$--algebra
is necessary) and select only contributions which potentially diverge as
$1/\e^2$. 
The effective diagrams after completion of $D$--algebra are shown in Fig. $4$
where, by suitable integrations by parts on the external sources, diagrams
on the first line are proportional to $\bar{D}^2J D^2 J$ and the ones on the
second line to $\bar{D}^{\dot{\a}} J D^{\a}J$. 

\vskip 18pt
\noindent
\begin{minipage}{\textwidth}
\begin{center}
\includegraphics[width=0.80\textwidth]{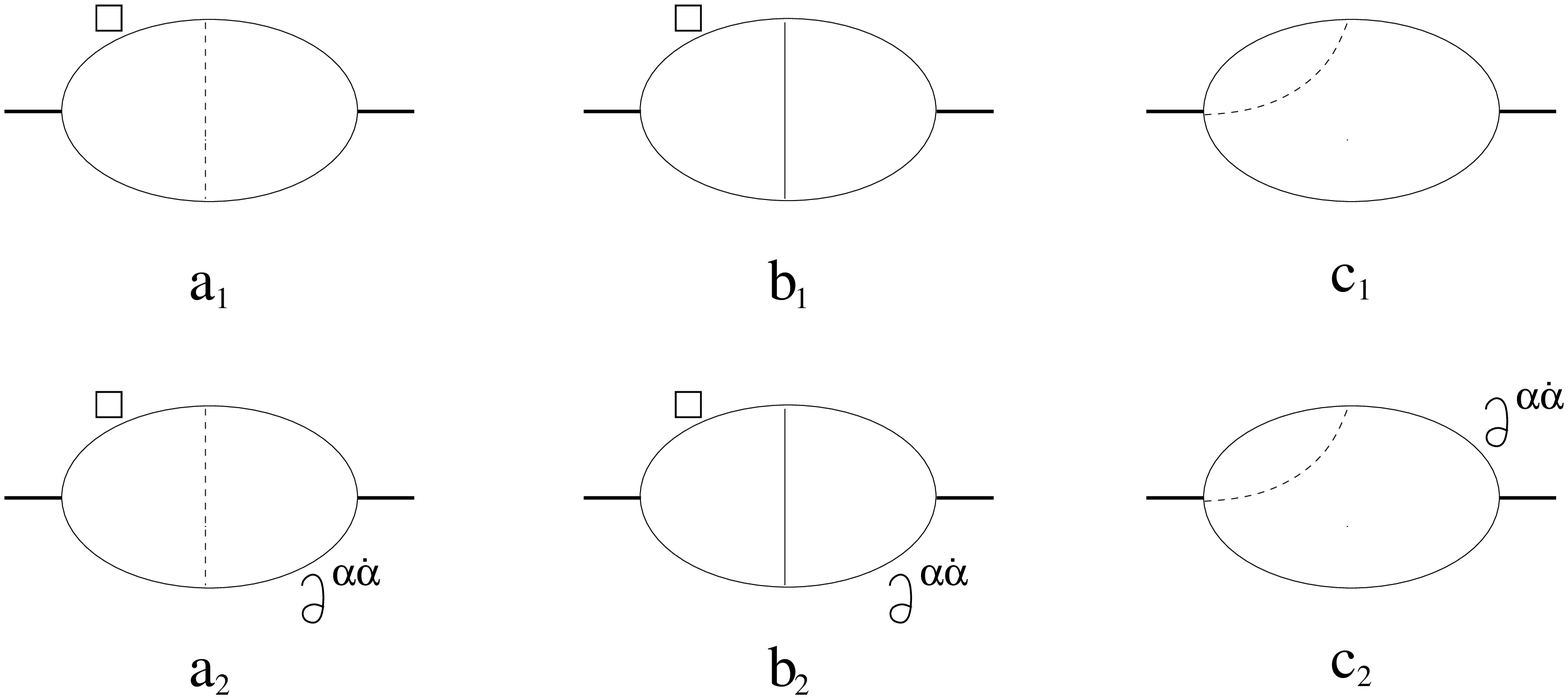}
\end{center}
\begin{center}
{\small{Figure 4: Diagrams from $(3a)-(3c)$ after completion of $D$--algebra}}
\end{center}
\end{minipage}

\vskip 20pt
\noindent
The first three diagrams reduce to the same momentum integral whose leading
divergence for $\e \to 0$ can be easily inferred by repeated use of 
identity (\ref{selfenergy2})
\beq
I ~\equiv~ \int \frac{d^n k d^n q}{ k^2 (p-k)^2 q^2 (q-k)^2} ~\sim~
\frac{1}{2 \e^2} 
\eeq
These diagrams contain a divergent subdiagram associated with
the $q$--loop. The subtraction of this subdivergence gives a
contribution $-\frac{1}{\e^2}$. Therefore, the actual leading divergence 
from these diagrams is 
\beq
I ~-~ I_{sub} ~\sim~ -\frac{1}{2\e^2}
\label{sub}
\eeq
The diagrams on the second line of Fig. $4$ give rise to the
momentum integral
\beq
\int \frac{d^n k d^n q ~k_{\a \dot{\a}}}{ k^2 (p-k)^2 q^2 (q-k)^2}
\eeq
Using identities (\ref{selfenergy2}, \ref{selfenergy3}), for $\e \to 0$  
its leading behavior
is given by $\frac12 p_{\a \dot{\a}} (I-I_{sub})$.

Evaluating the momentum integral associated with diagram $(3d)$ gives 
a simple pole divergence $1/\e$ (see (\ref{a2})). 
Therefore, this diagram
does not contribute to the anomalous dimension and we will
neglect it in what follows.
   
At this point we need only compute the overall factor from
combinatorics and group structures. 
Using identities in Appendix A, we find
\bea
&& (3a) ~:~ \,  ~~~~6 g^2 N( N^2 -1)
\nonumber \\
&& (3b) ~:~ \,  -12 g^2 N( N^2 -1)
\nonumber \\
&& (3c) ~:~ \,  -12 g^2 N( N^2 -1)
\ena
Restoring a factor $1/(4\pi)^2$ for each loop (see Appendix A)
the final leading contribution
at order $g^2$ to the generating functional is
\beq
W[J]_{1-loop} ~\rightarrow~ \frac{1}{\e^2} \, 9 \, \frac{g^2N}{(4\pi)^4} 
(N^2-1) 
\, \int \frac{d^4p}{(2\pi)^4} d^4 \th \,
J(-p, \th, \thb) \left[ \bar{D}^2 D^2 
~+~ \frac12 p_{\a \dot{\a}} \bar{D}^{\dot{\a}} D^{\a} \right] J(p, \th, \thb)  
\label{oneloop} 
\eeq
According to the general prescription given in Section 2, the anomalous
dimension at order $g^2$ is then given by the ratio of the coefficients
in (\ref{oneloop}) and (\ref{tree}), times $(\g_0-1)$. 
Taking into account that for the Konishi operator $\g_0=2$, we finally obtain  
the correct result $\g = \frac{3g^2N}{8\pi^2}$.

Before closing this Section, we would like to make some comments on the
connection between higher order poles in $\e$ and the appearance of a
nonvanishing anomalous dimension. From our calculation it is clear that
$1/\e^2$ poles can arise only from diagrams with self--energy 
subdiagrams containing {\em only} one external vertex (self--energy 
subdiagrams containing both, when subtracted, give rise to tadpole--like
integrals which vanish in dimensional regularization). On the other hand,
these subdiagrams are exactly the ones responsible for the renormalization
of the composite operator (in our particular example one can easily
convince by comparing the topologies of divergent subdiagrams in Fig. $4$
with the ones in Fig $1$ with $D$ algebra solved).  
This observation, toghether with the fact that multiple self--energy 
diagrams have only simple pole divergences (see eq. (\ref{a2})), 
gives an intuitive explanation of 
why in the evaluation of two--point functions higher order poles emerge 
only in the presence of anomalous dimensions.

\section{Anomalous dimension of the {\bf $O^{\bf (20)}_{ij}$} 
scalar superfield}

The gauge invariant $O^{\bf (20)}_{ij}$ superfield built in Appendix B 
(see eq. (\ref{O20}))
\beq
O^{\bf (20)}_{ij} ~\equiv~ {\rm Tr}(\Phi_k \Phi_{(i})
{\rm Tr}(e^{gV} \Phi_{j)} e^{-gV} \Phib^k -\frac{1}{3}\d_{j)}^k \,
e^{gV} \Phi_l e^{-gV} \Phib^l)
\label{O20bis}
\eeq
contains, as lowest component, a double trace scalar operator of dimension 
4 in the ${\bf 20}$ of $SU(4)$ which is expected not to renormalize 
perturbatively \cite{AFP1,AEPS,AES}.   
We apply our method to evaluate its anomalous dimension pertubatively
in superspace with the purpose to give an independent check of the previous
results.

To compute the two--point correlation function $\langle O^{\bf (20)}_{ij} 
(z_1) \bar{O}^{\bf (20)}_{i'j'} (z_2) \rangle$ 
we find convenient to separate the flavor 
contributions from the rest of the calculation. To this end we 
write the operator as
\beq
O^{\bf (20)}_{ij} ~=~ {\cal D}^{klpq}_{(ij)} ~ {\rm Tr} (\Phi_k \Phi_l) 
{\rm Tr} (e^{gV} \Phi_p  e^{-gV} \Phib_q) 
~\equiv~ {\cal D}^{klpq}_{(ij)} ~ O_{klpq}
\eeq
with the flavor tensor ${\cal D}$ given by
\beq
{\cal D}^{klpq}_{(ij)} ~=~ \frac12 \d^{q(k} \d^{l)(i} \d^{j)p} ~-~
\frac13 \d^{l(i} \d^{j)k} \d^{pq} 
\label{tensorD}
\eeq
It is symmetric in $(kl)$ and satisfies the property 
\beq
{\cal D}^{klpq}_{(ij)} \, \d_{pq} ~=~ 0
\label{property}
\eeq 
We then compute $\langle O_{klpq}(z_1) \bar{O}_{k'l'p'q'}(z_2)\rangle$ 
both at tree level and one--loop and leave the contructions with the
flavor tensors ${\cal D}^{klpq}_{(ij)}$ and 
${\cal D}^{k'l'p'q'}_{(i'j')}$ at the end.

The calculation follows the general recipe described in the previous Sections.
At tree level the two--point function is given by the diagram in Fig. $5$

\vskip 18pt
\noindent
\begin{minipage}{\textwidth}
\begin{center}
\includegraphics[width=0.60\textwidth]{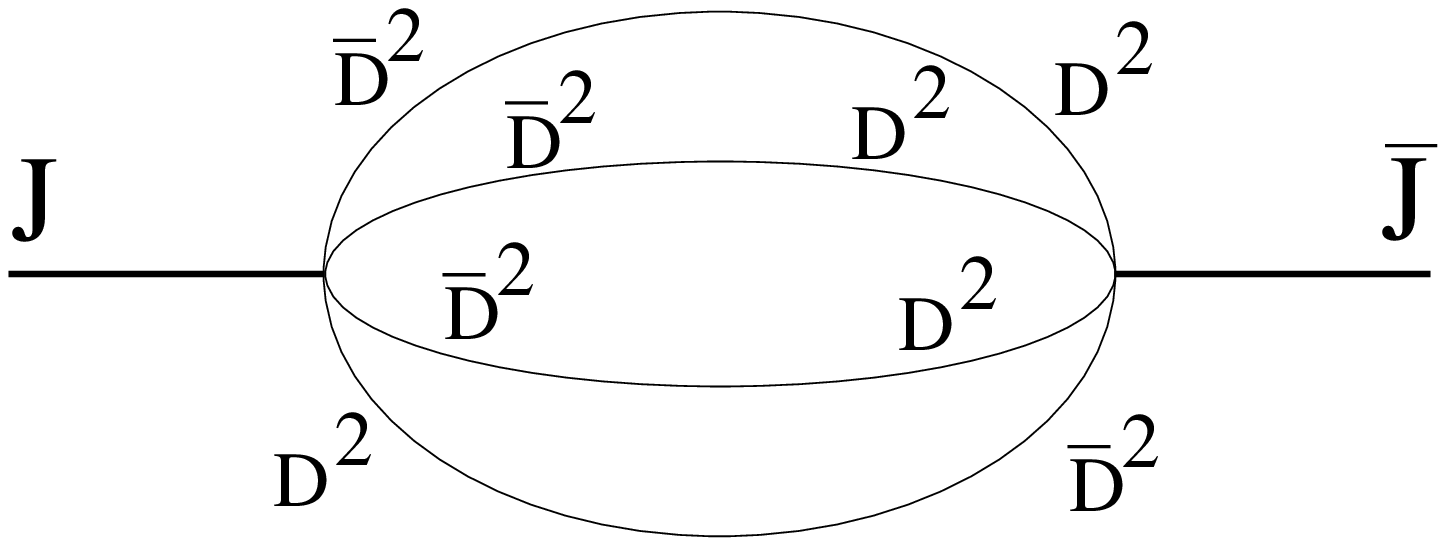}
\end{center}
\begin{center}
{\small{Figure 5: Tree level contribution to $\langle O^{\bf (20)}_{ij}(z_1) 
\bar{O}^{\bf (20)}_{i'j'}(z_2)\rangle$}}
\end{center}
\end{minipage}

\vskip 20pt

The D--algebra performed on this diagram is similar to the one for the 
Konishi superfield and eventually produces two independent 
superspace structures, $\bar{D}^2 J^{ij} D^2 \bar{J}^{i'j'}$ associated to a 
multiple self--energy integral
\beq
I ~\equiv ~
\int \frac{ d^nq_1 d^nq_2 d^nq_3}{q_1^2 (q_2 - q_1)^2 (q_3 - q_2)^2
(p - q_3)^2} 
\label{first}
\eeq
and $\bar{D}^{\dot{\a}} J^{ij} D^{\a} \bar{J}^{i'j'}$ associated to
\beq
I_{\a \dot{\a}} ~\equiv ~
\int 
\frac{ d^nq_1 d^nq_2 d^nq_3 ~(q_1)_{\a \dot{\a}}}{q_1^2 (q_2 - q_1)^2 
(q_3 - q_2)^2 (p - q_3)^2}  
\eeq
The momentum integral (\ref{first}) can be easily evaluated by
making use of (\ref{a2}) for $k=4$ and $a=0$. Its leading term is 
$\frac{1}{(3!)^2} \frac{1}{\e} (p^2)^2$. 
For the second integral, we use identity (\ref{a4}) 
to obtain $I_{\a \dot{\a}} \sim \frac{1}{4} p_{\a \dot{\a}} I$ 
for $\e \to 0$.
 
Performing the flavor and colour combinatorics and using the normalization
(\ref{norm}) one finds
\beq
\d^{bb'}_{qq'} ~\sum_{\s} \d^{a \s(a')}_{k \s(k')}
\, \d^{a \s(a')}_{l \s(l')} \, \d^{b \s(b')}_{p \s(p')} ~\equiv~ 
{\cal T}
\label{treecomb} 
\eeq
where $a,b,a',b'$ are colour indices and the short notation
$\d^{aa'}_{kk'} \equiv \d^{aa'} \d_{kk'}$ has been used. A sum over all
possible permutations of flavor and colour indices is explicitly indicated.
The colour combinatorics can be easily evaluated. Using the labellings 
(\ref{flavors}) for the different flavor structures 
the previous result can be written as
\beq
{\cal T} ~=~ (N^2-1)^2 \, \left( {\cal F}_1 ~+~ {\cal F}_4 \right)
~+ ~(N^2-1) \, \left( {\cal F}_2 ~+~ {\cal F}_3 
~+~ {\cal F}_5 ~+~ {\cal F}_6 \right)  
\eeq
As a last step we have to contract the previous expression with the external
tensors ${\cal D}^{klpq}_{(ij)}$ and ${\cal D}^{k'l'p'q'}_{(i'j')}$.
Making use of the identities (\ref{contractions}) we find 
\beq
{\cal D} {\cal T} {\cal D}' ~=~ \frac{20}{9} (N^2-1) (3N^2-2)
\, (\d_{ii'} \d_{jj'} ~+~ \d_{ij'} \d_{i'j} ) 
\label{T2}
\eeq
Therefore, reinserting a factor
$\frac{1}{(4\pi)^2}$ for each loop according to our conventions, 
we can finally write
\bea
\label{O20tree}
W[J,\bar{J}]_{tree} \rightarrow && \frac{1}{\e} \, \frac{1}{(3!)^2 (4\pi)^6}
\, \frac{20}{9} \, (N^2 - 1) \, (3N^2 -2)\\
&& ~~\times \, 2 \,
\int \frac{d^4p}{(2\pi)^4} d^4 \th \, (p^2)^2 \,
J^{ij}(-p,\th,\thb) \, \left[ \bar{D}^2 D^2 
~+~\frac{1}{4} \, p_{\a \dot{\a}}\bar{D}^{\dot{\a}} D^{\a} \right] \, 
\bar{J}^{ij}(p,\th,\thb)
\nonumber  
\eea

We now concentrate on the one--loop contribution.
All possible diagrams contributing at this order to the quadratic
terms in $W[J,\bar{J}]$ are listed in Fig. $6$.   
\noindent
Performing the $D$--algebra produces a huge amount of 
potentially divergent diagrams with different structures 
of spinorial derivatives acting on the external sources. However, 
using the results in (\ref{a2}, \ref{a4}, \ref{a3}) 
allows to conclude that most of the
diagrams have only simple pole divergences, so they can be neglected 
when computing anomalous dimensions. 
Moreover diagram $6g$ gives zero after summing over colour and flavor indices.
The only nonvanishing diagrams which manifest a 
$1/\e^2$ pole are the diagrams which reduce, after $D$--algebra, to the
topological structures in Fig. $7$. 
Diagrams $7a$, $7b$ and $7c$ are produced by  
$6a$, $6c$, $6d$ and $6e$ while $7d$ and $7e$ come from $6h$
\footnote{We are very grateful to  G.~Rossi and Y.~Stanev for having 
pointed us the nonvanishing contribution from diagram $6h$}.  

\vskip 18pt
\noindent
\begin{minipage}{\textwidth}
\begin{center}
\includegraphics[width=0.80\textwidth]{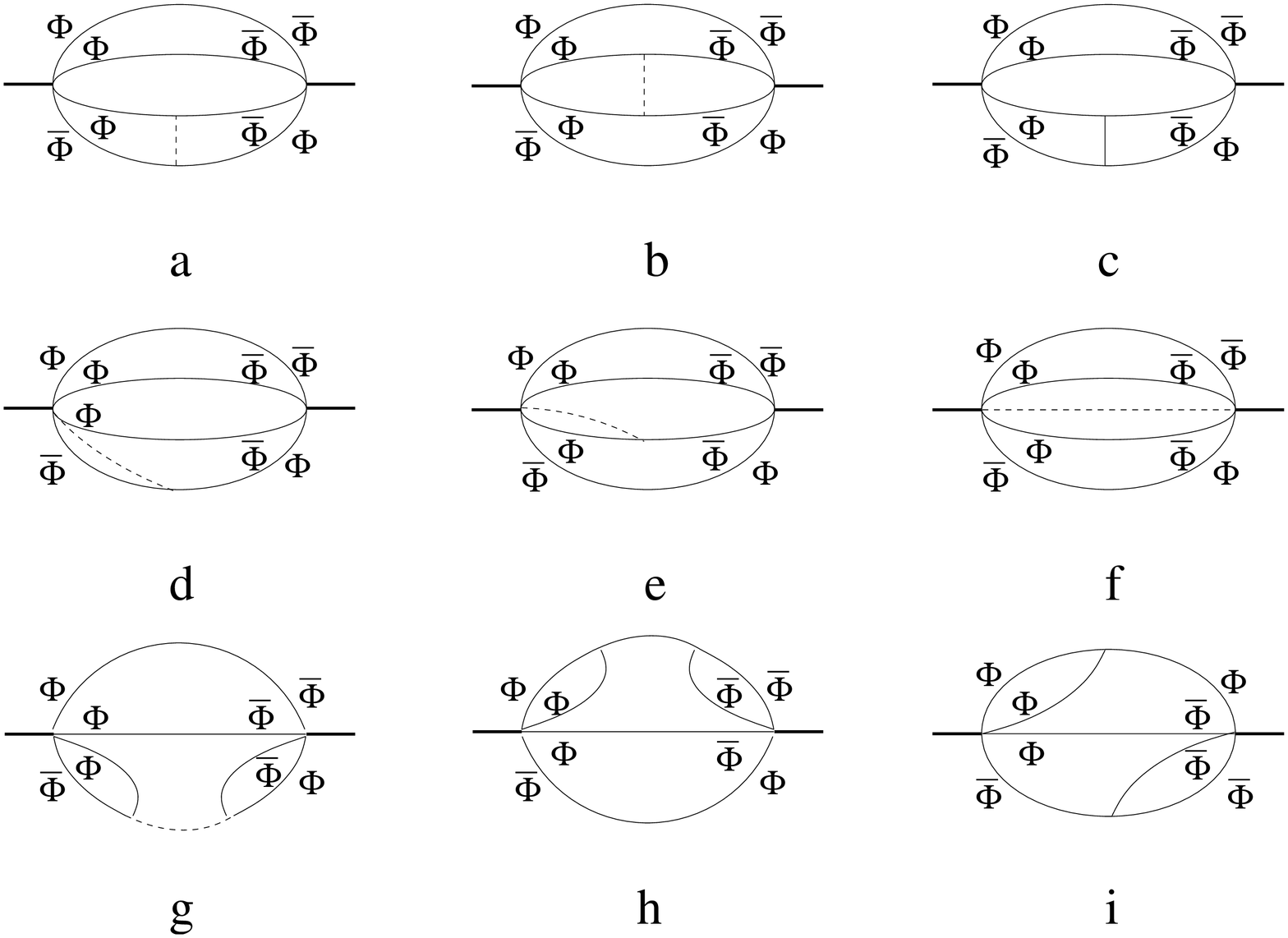}
\end{center}
\begin{center}
{\small{Figure 6: One--loop contributions to $\langle O^{\bf (20)}_{ij}(z_1)
\bar{O}^{\bf (20)}_{i'j'}(z_2)\rangle$}}
\end{center}
\end{minipage}

\vskip 20pt

\vskip 18pt
\noindent
\begin{minipage}{\textwidth}
\begin{center}
\includegraphics[width=0.80\textwidth]{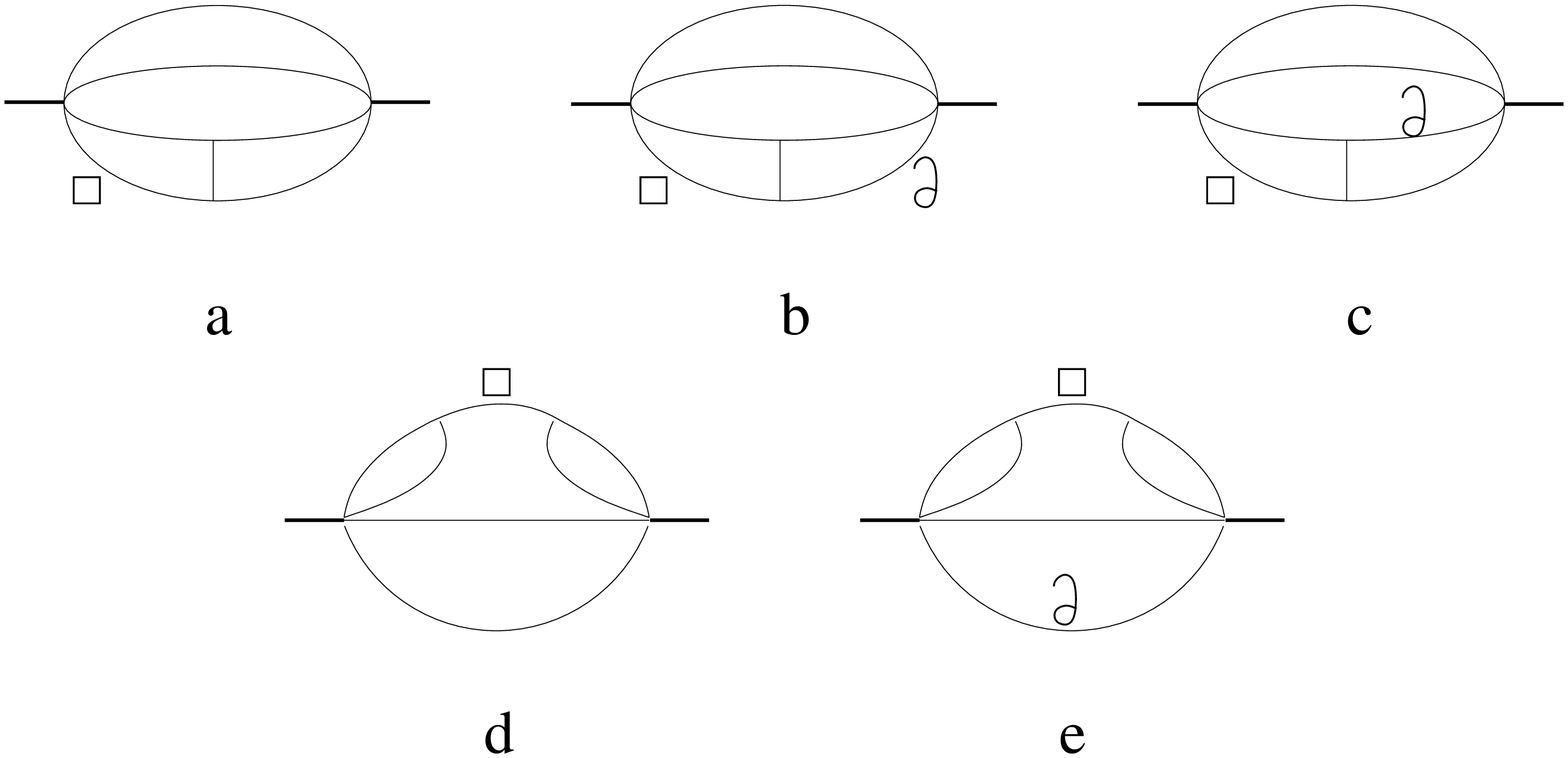}
\end{center}
\begin{center}
{\small{Figure 7: One--loop diagrams relevant for the anomalous dimension}}
\end{center}
\end{minipage}

\vskip 20pt

By suitable integration by parts 
the external field content of diagrams of type $7a$ and $7d$ can be written in 
the form $\bar{D}^2 J D^2 \bar{J}$, whereas diagrams in Figs. $7b$, $7c$ and
$7e$ 
are proportional to $\bar{D}^{\dot{\a}} J D^{\a} \bar{J}$. 
The corresponding integrals, with momentum lines as given in Fig. $8$, are
\bea
I^{7a} &~\equiv~& \int \frac{d^n q_1 \, d^n q_2 \, d^n q_3}{q_1^2 
(q_2 - q_1)^2 
(q_3- q_2)^2 (p - q_3)^2} ~ \int \frac{d^n k}{k^2 (q_1 - k)^2}
\label{int1}\\
&&~~~~~~~~~~~~~~\nonumber \\
I^{7b} &~\equiv~& \int \frac{d^n q_1 \, d^n q_2 \, d^n q_3 ~(p-q_3)_{\a \dot{\a}}}{q_1^2 
(q_2 - q_1)^2 (q_3- q_2)^2 (p - q_3)^2} ~ \int \frac{d^n k}{k^2 (q_1 - k)^2}
\nonumber\\
&&~~~~~~~~~~~~~~\nonumber \\
I^{7c} &~\equiv~& \int \frac{d^n q_1 \, d^n q_2 \, d^n q_3 ~(q_1)_{\a \dot{\a}}}{q_1^2 
(q_2 - q_1)^2 (q_3- q_2)^2 (p - q_3)^2} ~ \int \frac{d^n k}{k^2 (q_1 - k)^2}
\nonumber
\eea
and
\bea
I^{7d} &~\equiv~& \int \frac{d^n q_2 \, d^n q_3}{ 
(q_3- q_2)^2 (p - q_3)^2} ~ \int \frac{d^n k}{k^2 (q_2 - k)^2}
~\int \frac{d^n r}{r^2 (q_2 - r)^2}
\label{int2}\\
&&~~~~~~~~~~~~~~\nonumber \\
I^{7e} &~\equiv~& \int \frac{d^n q_2 \, d^n q_3 ~(p-q_3)_{\a \dot{\a}}}
{(q_3 - q_2)^2 (p - q_3)^2} ~ \int \frac{d^n k}{k^2 (q_2 - k)^2}
~\int \frac{d^n r}{r^2 (q_2 - r)^2}
\nonumber
\eea

\vskip 18pt
\noindent
\begin{minipage}{\textwidth}
\begin{center}
\includegraphics[width=0.70\textwidth]{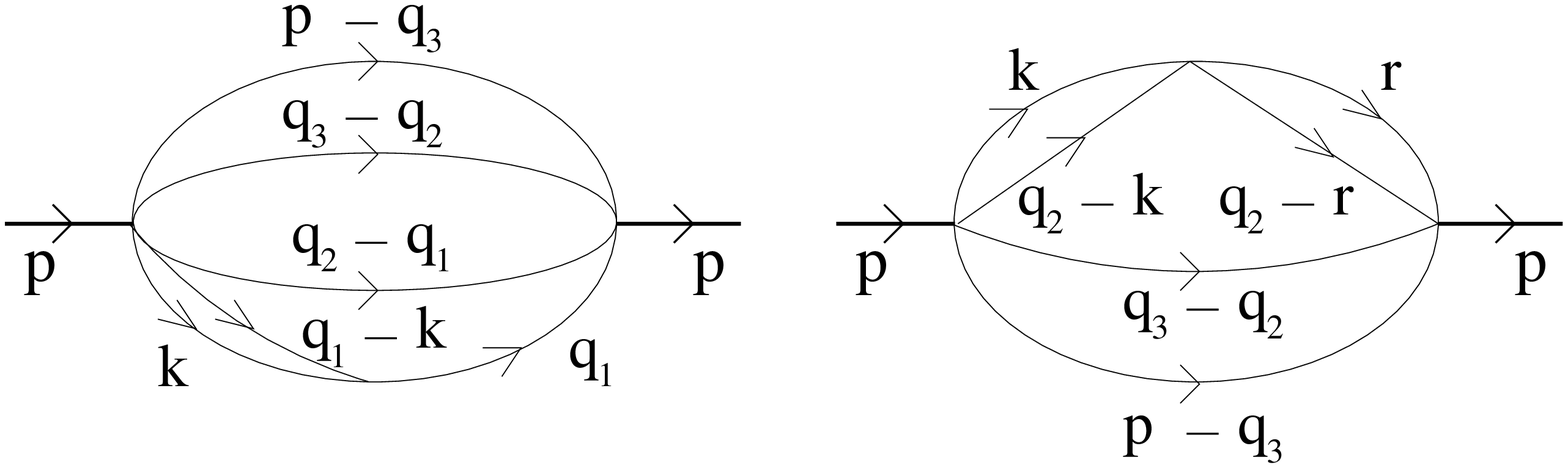}
\end{center}
\begin{center}
{\small{Figure 8: Momentum structures of leading divergent diagrams}}
\end{center}
\end{minipage}

\vskip 20pt

The $k$ and $r$ self--energy integrals are given in (\ref{selfenergy2}), 
while the rest
of the integrations can be easily performed with the help of identities
(\ref{a2},\ref{a4}).\\ 
For the first integral the result is 
\beq
I^{7a} ~\sim ~ 
\frac{1}{\e^2} \, \frac{1}{48} (p^2)^2 \quad , \quad 
\e\to 0
\eeq
The diagram contains a divergent subdiagram associated to the  
self--energy $k$ integral. Its subtraction corresponds to an extra 
contribution $-\frac{1}{\e^2} \, \frac{1}{36} (p^2)^2$. The final leading
term of the momentum integral (\ref{int1}) is then
\beq
I^{7a} ~-~ I^{7a}_{sub} ~\sim~ - \frac{1}{\e^2} \, \frac{1}{4 (3!)^2} (p^2)^2
\label{integral1}
\eeq
whereas the leading behavior of the integrals $I^{7b}$ and $I^{7c}$ is simply 
given by
$\frac{1}{4} p_{\a\dot{\a}} (I^{7a} - I^{7a}_{sub})$, so that the tree-level
structure $~\bar{D}^2 D^2 
+\frac{1}{4} \, p_{\a \dot{\a}}\bar{D}^{\dot{\a}} D^{\a}~$ is reproduced 
(see \ref{O20tree}).\\
Similarly, the integral (\ref{int2}) is given by 
\beq
I^{7d} ~\sim ~ 
\frac{1}{\e^2} \, \frac{1}{24} (p^2)^2 \quad , \quad 
\e\to 0
\eeq
and after the subtraction of the two $k$ and $r$ divergent subdiagrams
\beq
I^{7d} ~-~ I^{7d}_{sub} ~\sim~ - \frac{1}{\e^2} \, \frac{1}{2 (3!)^2} (p^2)^2
\label{integral2}
\eeq
whereas the leading behavior of the integral $I^{7e}$ is given by
$\frac{1}{4} p_{\a\dot{\a}} (I^{7d} - I^{7d}_{sub})$, so again the tree-level
structure $~\bar{D}^2 D^2 
+\frac{1}{4} \, p_{\a \dot{\a}}\bar{D}^{\dot{\a}} D^{\a}~$ is reproduced.

We are now left with the evaluation of the combinatorial factors from 
each diagram. The combinatorics from diagrams $6a$, $6d$ and $6e$
simply reduce to the tree level structure. 
Keeping also coefficients from vertices, propagators and combinatorics
from $D$--algebra we have
\bea
&& (6a): \qquad  ~~4g^2 N \, {\cal T} \nonumber \\
&& (6d): \qquad  -4g^2 N \, {\cal T} \nonumber \\
&& (6e): \qquad  -4g^2 N \, {\cal T}
\label{ADE} 
\eea
where the tensor ${\cal T}$ has been defined in (\ref{treecomb}).
The evaluation of the combinatorics for diagrams $6c$ and $6h$ is more 
involved. 
Indeed, in these cases the flavor indices from the internal vertices enter
the expression through the product of two Levi--Civita tensors 
(see eq. (\ref{vertices})) and the use of the standard identity 
\beq
\e_{ijk} \e_{imn} ~=~ \d_{jm} \d_{kn} ~-~ \d_{jn} \d_{km}  
\eeq
produces new flavor structures which do not appear at tree level.  
A long but straightforward calculation gives
\bea
(6c): \quad && ~~~4g^2 N \, {\cal T} \nonumber \\
&&~~~~~~~~~~~~~\nonumber \\
&& -~4g^2 N(N^2-1) ~\big[ 2 {\cal F}_1  - {\cal F}_2
-{\cal F}_3 - {\cal G}_1 -{\cal G}_2 
~+~ k \leftrightarrow l \big]   \nonumber \\
&&~~~~~~~~~~~~~\nonumber \\
&& -~ 4 g^2 N(N^2-1)^2 ~ \d_{pq} \d_{p'q'} ( \d_{ll'} \d_{kk'} ~+~
\d_{lk'} \d_{kl'} ) \nonumber \\
&&~~~~~~~~~~~~~\nonumber \\
&& -~ 4g^2 N(N^2-1) ~ \d_{pq} (\d_{k'q'} \d_{kp'} \d_{ll'} ~+~ 
\d_{l'q'} \d_{kk'} \d_{lp'}~+~
\d_{l'q'} \d_{kp'} \d_{lk'}~+~
\d_{k'q'} \d_{kl'} \d_{lp'}) \nonumber \\
&&~~~~~~~~~~~\nonumber \\
&& -~ 4g^2 N(N^2-1) ~ \d_{p'q'} (\d_{kq} \d_{pk'} \d_{ll'} ~+~ 
\d_{lq} \d_{kk'} \d_{pl'}~+~
\d_{lq} \d_{pk'} \d_{kl'}~+~
\d_{kq} \d_{lk'} \d_{pl'}) \nonumber \\
&&~~~~~~~~~~~ 
\label{partial}\\
&&~~~~~~~~~~~\nonumber\\
(6h): \quad && -~2g^2 N(N^2-1) ~\big[ 2 {\cal F}_1  - {\cal F}_2
-{\cal F}_3 ~+~ k \leftrightarrow l \big] 
\eea
At this point it is easy to perform the contractions with the flavor
tensors ${\cal D}$ and ${\cal D}'$. 
By exploiting the identity (\ref{property}) we immediately 
realize that the last three terms in (\ref{partial}) give vanishing 
contributions. By using the results 
in (\ref{contractions}, \ref{contractions2}) we finally have
\bea
&& (6a): \qquad ~~4 g^2 N \, {\cal D} {\cal T} {\cal D}'
\nonumber \\  
&& (6c): \qquad  ~~4g^2 N \, {\cal D} {\cal T} {\cal D}' 
~+~ 4g^2 N (N^2-1) \, \frac{100}{9} (\d_{ii'} \d_{jj'} + \d_{ij'} \d_{i'j})
\nonumber \\ 
&& (6d): \qquad -4 g^2 N \, {\cal D} {\cal T} {\cal D}'
\nonumber \\
&& (6e): \qquad -4 g^2 N \, {\cal D} {\cal T} {\cal D}' 
\nonumber\\
&& (6h): \qquad -2 g^2 N (N^2-1) \, \frac{100}{9} (\d_{ii'} \d_{jj'} + 
\d_{ij'} \d_{i'j})
\eea
where ${\cal D}{\cal T}{\cal D}'$ is given in (\ref{T2}). 

At this point we take into account also the factors from
the integrals, in particular we note 
that the integral (\ref{integral2}) coming from diagram $6h$ has an extra 
factor of 2 with respect to the integral (\ref{integral1}) coming from all 
the other diagrams. 
If we sum all the contributions we then have complete 
cancellation of the leading ${\cal T}$--terms, and also of the subleading 
terms produced by diagrams $6c$ and $6h$, so we can conclude
\beq
W[J,\bar{J}]_{one-loop} ~=~0
\eeq
proving that at order $g^2$ the anomalous dimension
of the double trace operator $O^{\bf (20)}_{ij}$ 
vanishes, in agreement with previous results \cite{AFP1,AEPS,AES}.

Our result gives further support to the belief
that the double trace superfield (\ref{O20bis}) is the 
protected operator whose existence is expected on the basis of general 
arguments \cite{AEPS,AES,HH}.

\section{Conclusions}

In this paper we have developed a rather simple procedure to compute 
anomalous dimensions directly from the two--point correlation functions.
Our prescription works in {\em any} superconformal field theory in four 
dimensions and allows to read perturbative contributions to unprotected
operators directly from few higher order poles divergent diagrams.  
We have applied this procedure to the particular case of ${\cal N}=4$ 
supersymmetric Yang--Mills theory with gauge group $SU(N)$, written in
${\cal N}=1$ superspace. 
Our result represents the first direct superspace calculation of anomalous
dimensions of operators other than Konishi \cite{AGJ,AFGJ}.\\
After a check
of the method on the anomalous dimension of the Konishi superfield, we have
evaluated at lowest order the anomalous dimension
of the double trace superfield $O^{\bf (20)}_{ij}$ in the ${\bf 20}$ of 
$SU(4)$. 
The result is that at one--loop $O^{\bf (20)}_{ij}$ does not 
renormalize. This is in agreement with previous 
calculations done in components, but has the advantage to keep
${\cal N}=1$ supersymmetry manifest.

The way we carried on the calculation for $O^{\bf (20)}_{ij}$ allows for
a quite immediate extension to other superfield representations, in particular
in the decomposition (\ref{20per20}). This subject will be extensively 
studied 
elsewhere \cite{us}.

\medskip

\section*{Acknowledgements}
\noindent
We acknowledge useful discussions with L.~Andrianopoli, M.~Pernici, 
A.C.~Petkou, G.~Rossi and Y.~Stanev. 
This work has been supported in 
part by INFN, MURST and the European Commission RTN program
HPRN--CT--2000--00131, in which S.P. is associated to
the University of Padova and A.S. is associated to the University of
Torino.

\newpage

\appendix

\section{Conventions and basic rules}

In this Appendix we list the main ingredients we need to compute perturbative
divergent contributions to two--point correlators as described in the main
text.
 
The classical action for ${\cal N}=4$ super Yang--Mills theory is given in
(\ref{actionYM}). Its quantization requires the introduction of a gauge fixing
and the corresponding ghosts.
The ghost superfields only couple to the vector
multiplet and do not enter our calculation at order $g^2$.
Working in Feynman gauge and momentum space the superfield propagators 
are
\beq
<V^a V^b>= - \frac{\d^{ab}}{p^2}\qquad\qquad
<\Phi^a_i \Phib^b_j>=\d_{ij} \frac{\d^{ab}}{p^2}
\label{propagators}
\eeq
The vertices can be obtained from the interaction terms in (\ref{actionYM}).
The ones that we need are the following
\bea
&&V_1=ig f_{abc}\d^{ij} \Phib^a_i V^b \Phi^c_j \nonumber \\
&& V_2 = - \frac{g}{3!} \e^{ijk} f_{abc} \Phi_i^a \Phi_j^b \Phi_k^c
\qquad \qquad \qquad
\bar{V}_2 = - \frac{g}{3!} \e^{ijk} f_{abc} \Phib_i^a \Phib_j^b \Phib_k^c
\nonumber \\
&&~~~~~~~~~~~
\label{vertices}
\eea
with additional $\Db^2$, $D^2$ factors for chiral, antichiral lines
respectively.

All the calculations are performed in $n$ dimensions, with $n=4-2\e$ 
and in momentum space. The momentum integrals can be computed by making
repeated use of the one loop results 
\beq
\int  \frac{d^nk}{(k^2)^a [(p-k)^2]^b} ~=~
\frac{\G(a+b-\frac{n}{2} )}{\G(a)\G(b)} \frac{\G(\frac{n}{2}-a)
\G(\frac{n}{2} -b)}
{\G(n-a-b)} \frac{1}{(p^2)^{a+b-\frac{n}{2} }}
\label{selfenergy2}
\eeq
\beq
\int  d^nk \, \frac{k_{\a \dot{\a}} }{(k^2)^a [(p-k)^2]^b} ~=~
\frac12 \, \frac{n-2a}{n-a-b} p_{\a \dot{\a}} \, 
\int  \frac{d^nk}{(k^2)^a [(p-k)^2]^b}
\label{selfenergy3}
\eeq
In particular, the basic formulae are
\bea
I &~\equiv~& \int \frac{d^n q_1 \cdots d^n q_{k-1}}{(q_1^2)^{1+ a \e} 
(q_2-q_1)^2 (q_3-q_2)^2
\cdots (p-q_{k-1})^2} ~= \nonumber \\
&=& 
\frac{1}{(a+1)\e}\int \frac{d^n q_2 \cdots d^n q_{k-1}}{(q_2^2)^{(a+1) \e} 
(q_3-q_2)^2
\cdots (p-q_{k-1})^2} ~+~ {\cal O}(\e^0)~= \nonumber \\
&&~~~~~~~~~~~~~~~~~~~~~~~~~~~~~~~~~~~
\frac{1}{\e} \, \frac{(-1)^k \, (k-1)}{[(k-1)!]^2
\, (k-1+a)} \, (p^2)^{k-2-(k-1+a)\e} ~+~ {\cal O}(\e^0) \nonumber \\
&&~~~~~~~~~
\label{a2}
\eea
and
\bea
\int 
\frac{d^n q_1 \cdots d^n q_{k-1} ~ q_1^{\a \dot{\a}} }{(q_1^2)^{1+ a \e} 
(q_2-q_1)^2 (q_3-q_2)^2 \cdots (p-q_{k-1})^2} ~&\sim &~
\frac{1}{k} \, p^{\a \dot{\a}} ~I \nonumber\\
&&~~~\nonumber\\
\int 
\frac{d^n q_1 \cdots d^n q_{k-1} ~ (p-q_{k-1})^{\a \dot{\a}} }{(q_1^2)^{1+ a \e} 
(q_2-q_1)^2 (q_3-q_2)^2 \cdots (p-q_{k-1})^2} ~=~~~~~~~~~~&&\nonumber\\
&&~~~\nonumber\\
=~\frac{1}{(a+1)\e}
\int
\frac{d^n q_2 \cdots d^n q_{k-1} ~ 
(p-q_{k-1})^{\a \dot{\a}}}{ 
(q_2^2)^{(a+1)\e} (q_3-q_2)^2 \cdots (p-q_{k-1})^2} ~+~ {\cal O}(\e^0)
~&\sim &~
\frac{1}{k} \, p^{\a \dot{\a}} ~I 
\label{a4}
\eea
where in the last two equations the leading behavior for $\e \to 0$ has been
extracted.
Moreover, it is useful to remind the finiteness of the following two--loop
integral
\beq
\int \frac{d^nk~d^nl}{k^2l^2(k-l)^2(p-k)^2(p-l)^2}
~=~ \frac{1}{(p^2)^{1+2\e}}
[6\z(3)+{\cal O}(\e)]
\label{a3}
\eeq
Notice that we always neglect $2\pi$ factors in the momentum integrals 
with the convention that we reinstate a $1/(4\pi)^2$ factor for each loop
in the final result. 

Our basic conventions to deal with colour structures are the following.
For a general simple Lie algebra we have
\beq
[T_a, T_b] ~=~ i f_{abc} T_c
\label{algebra}
\eeq
where $T_a$ are the generators  and
$f_{abc}$ the structure constants. The matrices $T_a$'s are normalized as
\beq
{\rm Tr}(T_a T_b) ~=~\d_{ab}
\label{norm}
\eeq
We specialize to the case of $SU(N)$ Lie algebra whose generators $T_a$,
$a= 1, \cdots, N^2-1$ are taken in the fundamental represetation, i.e.
they are $N \times N$ traceless matrices.
The basic relation which allows to deal with products of $T_a$'s is the
following
\beq
T^a_{ij} T^a_{kl}= \left( \d_{il}\d_{jk}-
\frac{1}{N}\d_{ij}\d_{kl}\right).
\label{Tcontract}
\eeq
From this identity, toghether with (\ref{algebra}), we can easily obtain 
all the identities used to compute the colour 
structures associated to the Feynman diagrams relevant for the 
two--point correlation functions. They are 
\beq
f_{acd} f_{bcd} ~=~ 2N \, \d_{ab}
\eeq
\beq
{\rm Tr}(T_a T_b T_c T_d)~{\rm Tr}(T_a T_b T_c T_d) =  \frac{1}{N^2}
(N^2-1)(N^2+3)
\label{tr1}
\eeq
\beq
{\rm Tr}(T_a T_b T_c T_d)~{\rm Tr}(T_a T_b T_d T_c) = - \frac{1}{N^2}
(N^2-1)(N^2-3)
\label{tr2}
\eeq
\beq
{\rm Tr}(T_a T_b T_c T_d)~{\rm Tr}(T_d T_c T_b T_a) = \frac{1}{N^2}
(N^2-1)(N^4-3N^2+3)
\label{tr3}
\eeq
and
\beq
{\rm Tr}(T_{c_1} T_{a_1} T_{c_2} T_{a_2}) f_{c_1mb_1} f_{c_2mb_2} =
- (\d_{b_1a_1}\d_{b_2a_2} + \d_{b_2a_1}\d_{b_1a_2})
\label{tr4}
\eeq
\beq
{\rm Tr}(T_{c_1} T_{c_2} T_{a_1} T_{a_2}) f_{c_1mb_1} f_{c_2mb_2} =
~\d_{b_1b_2}\d_{a_1a_2} ~+~ N \, {\rm Tr}(T_{b_1} T_{b_2} T_{a_1} T_{a_2})
\label{tr5}
\eeq

To deal with the flavor structure we find convenient to introduce the 
following tensors (to shorten the notation
we avoid to write explicitly the flavor indices on the tensors)
\bea
&& {\cal F}_1 ~\equiv~ 
\d_{kk'} \, \d_{ll'} \, \d_{pp'} \, \d_{qq'} 
\qquad \qquad 
{\cal F}_4 ~\equiv~ 
\d_{kl'} \, \d_{lk'} \, \d_{pp'} \, \d_{qq'}
\nonumber \\
&& {\cal F}_2 ~\equiv~ 
\d_{kp'} \, \d_{lk'} \, \d_{pl'} \, \d_{qq'} 
\qquad \qquad
{\cal F}_5 ~\equiv~ 
\d_{kk'} \, \d_{lp'} \, \d_{pl'} \, \d_{qq'}
\nonumber \\ 
&& {\cal F}_3 ~\equiv~ 
\d_{kl'} \, \d_{lp'} \, \d_{pk'} \, \d_{qq'} 
\qquad \qquad
{\cal F}_6 ~\equiv~ 
\d_{kp'} \, \d_{ll'} \, \d_{pk'} \, \d_{qq'}
\label{flavors}
\eea
and 
\beq
{\cal G}_1 ~\equiv~ \d_{kq} \d_{k'q'} \d_{ll'} \d_{pp'} 
\qquad \qquad
{\cal G}_2 ~\equiv~ \d_{kq} \d_{l'q'} \d_{lk'} \d_{pp'}
\label{flavors2}
\eeq
The contractions with the tensors ${\cal D}^{klpq}_{(ij)} \equiv {\cal D}$ 
and ${\cal D}^{k'l'p'q'}_{(i'j')} \equiv {\cal D}'$ as defined in the
main text, eq. (\ref{tensorD}), give the following results
\bea
&& {\cal D} {\cal F}_1 {\cal D}' ~=~ {\cal D} {\cal F}_4 {\cal D}' ~=~ 
\frac{10}{3} ( \d_{ii'} \d_{jj'}
~+~ \d_{ij'} \d_{i'j} ) \nonumber \\
&& {\cal D} {\cal F}_2 {\cal D}' ~=~ {\cal D} {\cal F}_3 {\cal D}' ~=~ 
{\cal D} {\cal F}_5 {\cal D}' ~=~ 
{\cal D} {\cal F}_6 {\cal D}' ~=~ \frac{5}{9} ( \d_{ii'} \d_{jj'}
~+~ \d_{ij'} \d_{i'j} ) 
\label{contractions}
\eea  
and
\beq
{\cal D} {\cal G}_1 {\cal D}' ~=~ {\cal D} {\cal G}_2 {\cal D}' ~=~ 
\frac{50}{9} ( \d_{ii'} \d_{jj'}
~+~ \d_{ij'} \d_{i'j} ) 
\label{contractions2}
\eeq

\section{Construction of the $O^{\bf (20)}_{ij}$ superfield}

In this Appendix we describe in detail the procedure to construct 
in terms of $N=1$ superfields 
the dimension 4 operator in the ${\bf 20}$ of $SU(4)$ appearing in the product
of two dimension 2 CPO's, also belonging to the ${\bf 20}$.

The ${\cal N}=1$ fundamental superfields of ${\cal N}=4$ supersymmetric
Yang--Mills theory are chiral $\Phi_i$ and antichiral $\Phib^i$ superfields
($i=1,\ldots,3$) in the ${\bf 3}$ and ${\bf \bar{3}}$ of $SU(3)$, respectively.
The six scalars of the theory which belong to the (real) ${\bf 6}$ of the 
R--symmetry group $SO(6) \sim SU(4)$ can be expressed in terms of the lowest
components of these superfields according to the following decomposition 
under $SU(4)\rightarrow SU(3)\times U(1)$
\beq
{\bf 6}\rightarrow {\bf 3}_{(\frac{2}{3})} + {\bf \bar{3}}_{(-\frac{2}{3})}
\label{seidecomp}
\eeq
where the {\bf 3} and ${\bf \bar{3}}$ representations contain, respectively, 
the highest and the lowest weight of the {\bf 6}. The small indices
$(\frac{2}{3})$ and $(-\frac{2}{3})$ are (conventional) $U(1)$ charges.
\\ 
The decomposition (\ref{seidecomp}) can be deduced from the one of 
the fundamental {\bf 4} representation
of $SU(4)$
\beq
{\bf 4}\rightarrow {\bf 3}_{(-\frac{1}{3})} + {\bf 1}_{(1)}
\eeq
\\
Analogously, the decomposition of the (real) {\bf 20} can be found to be
\beq
{\bf 20}\rightarrow {\bf 6}_{(\frac{4}{3})} + {\bf 8}_{(0)} + 
{\bf \bar{6}}_{(-\frac{4}{3})}
\label{20decomp}
\eeq
where the {\bf 6} and ${\bf \bar{6}}$ representations contain, respectively, 
the highest and the lowest weight of the {\bf 20}.\\
We want to build the superfield expression corresponding to the highest 
weight {\bf 6} (or to its complex conjugate lowest weight ${\bf \bar{6}}$) 
that appear in the decomposition under (\ref{20decomp}) of our 
{\bf 20} operator on the right hand side of 
\beq
({\bf 20}\times{\bf 20})_S ={\bf 105}+{\bf 84}+{\bf 20}+{\bf 1}
\label{20per20}
\eeq 
To this purpose we use (\ref{20decomp}) on the left hand side of 
(\ref{20per20}) and we produce all the following terms
\bea
({\bf 6}_{(\frac{4}{3})}\times{\bf 6}_{(\frac{4}{3})})_S &=& 
{\bf \bar{6}}_{(\frac{8}{3})} + {\bf 15}_{(\frac{8}{3})} 
\nonumber\\
({\bf 8}_{(0)}\times{\bf 8}_{(0)})_S &=& {\bf 1}_{(0)} + {\bf 8}_{(0)} +
{\bf 27}_{(0)}
\nonumber\\
({\bf \bar{6}}_{(-\frac{4}{3})}\times{\bf \bar{6}}_{(-\frac{4}{3})})_S &=&
{\bf 6}_{(-\frac{8}{3})} + {\bf \bar{15}}_{(-\frac{8}{3})} 
\nonumber\\
{\bf 6}_{(\frac{4}{3})}\times{\bf 8}_{(0)} &=& 
{\bf \bar{3}}_{(\frac{4}{3})} + {\bf 6}_{(\frac{4}{3})} +
{\bf \bar{15}}_{(\frac{4}{3})} + {\bf 24}_{(\frac{4}{3})} 
\nonumber\\
{\bf 6}_{(\frac{4}{3})}\times{\bf \bar{6}}_{(-\frac{4}{3})} &=&
{\bf 1}_{(0)} + {\bf 8}_{(0)} + {\bf 27}_{(0)}
\nonumber\\
{\bf \bar{6}}_{(-\frac{4}{3})}\times{\bf 8}_{(0)} &=& 
{\bf 3}_{(-\frac{4}{3})} + {\bf \bar{6}}_{(-\frac{4}{3})} +
{\bf 15}_{(-\frac{4}{3})} + {\bf \bar{24}}_{(-\frac{4}{3})}
\eea
We see that the {\bf 6} can be obtained in two ways, from 
$({\bf \bar{6}}\times{\bf \bar{6}})_S$ and from ${\bf 6}\times{\bf 8}$.
However, only the last one has the correct $U(1)$ charge $+\frac{4}{3}$ 
(see eq. (\ref{20decomp})).\\
In terms of ${\cal N}=1$ superfields, objects in the {\bf 6} and {\bf 8} are 
obtained as bilinear forms (being $\g_0=2$) 
\beq
{\rm Tr}(\Phi_i \Phi_j) ~\equiv~ A_{ij}
\eeq
and 
\beq
{\rm Tr}(\Phi_i \Phib^j -
\frac{1}{3}\d_i^j \, \Phi_k \Phib^k) ~\equiv~ B_i^j
\eeq  
respectively. The operator in the ${\bf 6}\times{\bf 8}$ is then $A_{ij}B_k^l$.
The contraction of one of the chiral indices of the operator in the {\bf 6} 
with the antichiral index of the operator in the {\bf 8} gives the two 
irreducible representations
\beq
A_{k[i}B_{j]}^k ~~~~~~~~~ A_{k(i}B_{j)}^k
\eeq
where the first one is the ${\bf \bar{3}}$ and the second one is the
{\bf 6} appearing in the product ${\bf 6}\times{\bf 8}$.
The explicit form of the operator is then given by
\beq
{\rm Tr}(\Phi_k \Phi_{(i})
{\rm Tr}(\Phi_{j)} \Phib^k -\frac{1}{3}\d_{j)}^k \,
\Phi_l \Phib^l)
\label{O20senzaV}
\eeq
From the point of view of the $R$--symmetry group representations this is
the correct superfield structure we were looking for. However, 
in order to guarantee also its gauge invariance, we need introduce
exponentials of the gauge superfield $V$, obtaining
\beq
{\rm Tr}(\Phi_k \Phi_{(i})
{\rm Tr}(e^{gV} \Phi_{j)} e^{-gV} \Phib^k -\frac{1}{3}\d_{j)}^k \,
e^{gV} \Phi_l e^{-gV} \Phib^l)
\label{O20}
\eeq
This is the form of the superfield $O^{\bf (20)}_{ij}$ operator we have 
used in our calculation.

\end{document}